\documentclass[12pt,a4paper]{article}

\usepackage{epsf}
\usepackage{cite}
\begin{document}
\baselineskip 24pt
\title{ 
Long-time behavior and relaxation of power-law correlation 
in one-dimensional self-gravitating system
}

\author{
  Hiroko KOYAMA\thanks{\tt hiroko@allegro.phys.nagoya-u.ac.jp} \, 
and Tetsuro KONISHI\thanks{\tt tkonishi@allegro.phys.nagoya-u.ac.jp} \\
Department of Physics, School of Science,\\
Nagoya University, 464-8602, Nagoya, Japan
}
\date{}
\maketitle
  \begin{abstract}
Long-time behavior of spatial power-law correlation in one-dimensional 
self-gravitating system is numerically investigated. % It is found that 
The power-law structure persists even after the system is virialized.
The structure gradually disappears with energy exchange among 
particles. 
Lifetime of the power-law structure is estimated to be proportional to
the system size.

  \end{abstract}
\section{Introduction}
Spatial structure in many-body systems is an important subject
in many branches of physics, such as molecular systems, gravitational
systems, and so on.  
For various types of such structures, their origins are interesting subject 
of study.
Some of them 
may be realized as equilibrium distribution, and others may be 
already set at initial conditions.

Recently we have discovered that spatial structure with power-law correlation
spontaneously emerges from uniformly random initial conditions
in one-dimensional self-gravitating system~\cite{hktk-1}.
What is important in this novel phenomenon is that the spatial structure is not
given at the initial condition, but dynamically  created from a state
without spatial  correlation.
Succeeding research clarified that the structure is created first in 
small spatial scale then grows up to large scale
through hierarchical clustering~\cite{hktk-2}.

The question to be discussed now is the properties of the state 
which has power-law correlation. 
In particular we have to clarify whether the state persists for long time,
or just a short transient and disappears quickly.
If the state lasts for long time, we have good chance to observe it
and it has much importance. On the other hand, if the state 
disappears quickly, it may not be observable easily and it is  less 
important to the whole dynamics of the system.

Another important question is  on the relaxation process~( See 
\cite{Miller-2,sheet-tkg-1,sheet-tgk-2} and references therein.).
Since the system is an isolated many-body system without dissipation,
one would expect that
the system will eventually relax to thermal equilibrium for sufficiently
long time scale. 
Hence  relation between
decay (if it does) of power-law structure and 
relaxation processes of thermalization 
is also important.

The purpose of this paper is to investigate the long-time behavior of 
the power-law structure; its lifetime and mechanism of relaxation.
The model is introduced in the next section. In section \ref{sec:results}
we introduce several quantities which represent relaxation 
and investigate their dynamics to clarify the mechanism 
by which the power-law structure disappears.
The final section is devoted to summary and discussions.

\section{Model}
The model we use in this paper is the 
self-gravitating one-dimensional 
system~\cite{Miller-2,sheet-tkg-1,sheet-tgk-2,hohl-feix-1967,rybivki-1971-apss,severne-1984-apss,yamashiro-gouda-sakagami}.  
It represents a system of particles on one-dimension interacting 
through Newtonian gravity. It can also be interpreted as a system
consists of equivalent mass-sheets placed parallel in 3-dimensional space,
hence it is also called as the sheet model.

The Hamiltonian of the model is
\begin{eqnarray}
  H &=& \sum_{i=1}^N \frac{p_i^2}{2m} + 2\pi G m^2\sum_{i > j} \left|
x_i - x_j\right| \  , \ \ -\infty < x_i < \infty \ ,
  \label{eq:sheet-hamiltonian} 
\end{eqnarray}
where $x_i$ and $p_i$ represent coordinate and momentum of $i$'th particle,
$N$ is the number of particles,   $m$ stands for mass of each 
particle, and $G$ is a positive constant.
Force between two particles is attractive and long-ranged.

In this paper we set
$ m \equiv 1/N$ and $4\pi G \equiv 1$ \  .
Time is measured in unit
$t_c \equiv 1/\sqrt{4\pi G M/L}$ ,
where $M \equiv mN$ is total mass, $L$ is the spatial length 
on which particles are distributed initially. 

\section{Numerical results}\label{sec:results}
Now  we show temporal evolution of the 
power-law correlation. Since the formation process is already 
studied in our previous work~\cite{hktk-2}, here we focus on the long-time
behavior.

In what follows we  use a typical example 
whose initial condition and 
parameters are defined as follows: 
\begin{eqnarray}
  \label{eq:parameters}
\mbox{system size} & :& N=2048 \ , \nonumber\\
\mbox{initial condition} & :&  \forall i \ v_i = 0 , \\
 &{}& x_i = \mbox{uniformly random in }\ [0,1) .\nonumber% , \\
\end{eqnarray}

\subsection{Power-law correlation}

In our previous works we showed that spatial structure where
two-point correlation function obey power-law  
\begin{equation}
  \label{eq:power-law}
 \xi(r)\propto r^{-\alpha}
\end{equation}
is formed in the model~(\ref{eq:sheet-hamiltonian})
from uniformly-random initial conditions~\cite{hktk-1,hktk-2}.
Here the  the two-point correlation function $\xi(r)$ is defined as
\begin{equation}
dP = n dV (1 + \xi(r))
  \label{eq:2-body-correlation}
\end{equation}
where $dP$ stands for probability to find another particle 
in volume $dV$ at distance $r$ from a particle. $n$ is the 
average number density. 

\begin{figure}[hbtp]
    \epsfxsize=6cm
    \epsfbox{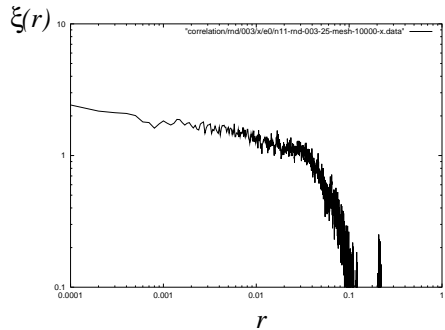}
    \caption{An example of power-law behavior of correlation function
      (\protect\ref{eq:2-body-correlation}) for parameters~(\protect
      \ref{eq:parameters}). $t=250.$ \       $\xi(r)\propto r^{-0.11}$.}
    \label{fig:corr-and-xu-rnd-003}
  \end{figure}

Using this correlation function we found that the 
model~(\ref{eq:sheet-hamiltonian}) with initial 
conditions with zero velocity dispersion evolves into a state where
$\xi(r)$ obeys power-law~(\ref{eq:power-law}).
 In Figure~\ref{fig:corr-and-xu-rnd-003} we show a typical 
example of such power-law correlation. 

\subsection{$V_r$  : virial ratio}

The main question we would like to discuss in this paper is 
whether the power-law structure lasts for long time or not.
For that purpose we need to define a time scale for reference.

As for the one-dimensional self-gravitating system~(\ref{eq:sheet-hamiltonian}),
it is known that relaxation proceeds through three stages;
first virialization process occurs, then ``microscopic relaxation'', then
``macroscopic relaxation''~\cite{sheet-tkg-1,sheet-tgk-2}.
(Each process is explained later.) We will compare the decay of power-law 
structure and each relaxation process.

First we adopt a time scale called virial time $t_{vr}$ 
defined from virial ratio 
$V_r(t)$. Virial ratio is  defined as
\begin{equation}
  \label{eq:vr}
  V_r(t)\equiv  2E_{kin}(t)/E_{pot}(t) \ \ ,
\end{equation}
where $\displaystyle E_{kin}(t)$ and $\displaystyle E_{pot}(t)$
are kinetic and potential energy of the system at time $t$, respectively.

Since the total energy is conserved and finite, 
it follows from virial theorem that 
\begin{equation}
  \label{eq:virial-theorem}
    2\overline{E_{kin}} /\overline{E_{pot}}  = 1 \ ,
\end{equation}
where 
$\displaystyle \overline{E_{kin}}$ and $\displaystyle \overline{E_{pot}}$
are long time average of 
$\displaystyle E_{kin}(t)$ and $\displaystyle E_{pot}(t)$, respectively.
Thus convergence of $V_r(t)$ to 1 is one good measure of relaxation
of the system, and the time scale $t_{vr}$ at which $V_r(t)$ converges
 to 1  is a good reference timescale. We call $t_{vr}$ 
virial time.

\begin{figure}[hbtp]
    \epsfxsize=6cm
    \epsfbox{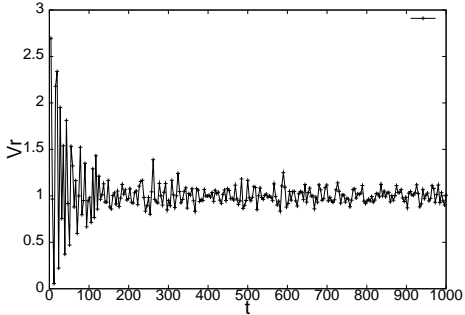}
    \caption{Time evolution of virial ratio $V_r(t)$(\protect\ref{eq:vr}) for parameters (\protect\ref{eq:parameters}). }
    \label{fig:vr}
\end{figure}

Figure~\ref{fig:vr} shows temporal evolution of the virial 
ratio~(\ref{eq:vr}).
It is seen that the virial ratio $V_r(t)$ converges to 1
at about  $t \sim 150$. At this time it can be said that the system has 
relaxed to a state in the sense of energy partition.
The important point to note is that the
power-law correlation shown in Fig.\ref{fig:corr-and-xu-rnd-003}
exists after the virial ratio is converged to 1.
(Remember that the figure is taken at $t=250$. Also it is confirmed 
in other examples.)
That is,  power-law correlation lasts longer than virial time.
In this sense the power-law correlation persists long.

\subsection{$\alpha(t)$  : exponent of power-law}

The next problem is the length of time it lasts.
For that purpose  we investigate 
the long-time behavior of the exponent $\alpha$
of power-law~(\ref{eq:power-law}).

\begin{figure}[hbtp]
      \epsfxsize=6cm
      \epsfbox{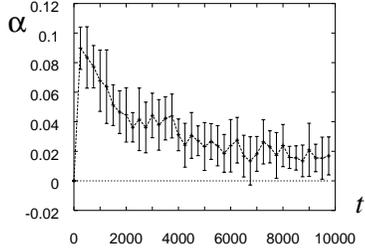}
   \caption{Long-time behavior of the exponent $\alpha(t)$ of
     correlation (\protect\ref{eq:power-law}) for $0 \le t \le 10000$ 
     for 10 initial conditions.}
   \label{fig:exponent-t}
 \end{figure}

Fig.~\ref{fig:exponent-t} represents temporal evolution of 
the exponent $\alpha$ for time scale much longer than 
the one  shown in Fig.~\ref{fig:vr} for 10 independent orbits.
Initial conditions are the same as (\ref{eq:parameters}) with 
different realization of random numbers. 
From this figure we see that the power-law structure is not stationary
but transient and gradually fading in this time scale. 
Decrease of the exponent $\alpha(t)$ can be considered as relaxation process.

\subsection{$\Delta(t)$  : energy exchange}
Next we investigate the process of relaxation of power-law correlation.
For that purpose we measure $\Delta(t)$ defined below,
by which we can see  the degree of energy exchange
between particles~\cite{sheet-tkg-1,sheet-tgk-2}.

$\Delta(t)$ is defined as follows. First we define $\varepsilon_i(t)$, 
1-particle energy per unit mass as
\begin{equation}
  \label{eq:energy-1}
\varepsilon_i(t) \equiv \frac{1}{2}v_i^2(t) 
+2\pi Gm \sum_{j=1}^N \left|x_j(t) - x_i(t) \right| \ \ .
\end{equation}

\begin{figure}
      \epsfxsize=6cm
      \epsfbox{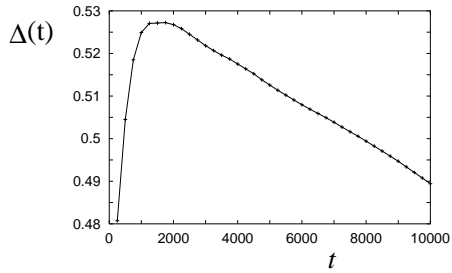}
      \caption{Temporal evolution of  
        $\Delta(t)$   for $0 \le t \le 10000$ .}
      \label{fig:delta}
  \end{figure}

If the system is ergodic the long time average of $\varepsilon_i(t)$ 
converges to a unique value 
\begin{equation}
  \lim_{T\rightarrow\infty}\frac{1}{T}\int_0^T \varepsilon_i(t)dt
= 5E_{total}/3 \equiv \varepsilon_0 
\end{equation}
for all $i$.

Using this $\varepsilon_0$ , $\Delta(t)$ is defined as
\begin{equation}
  \label{eq:delta}
  \Delta(t) \equiv \frac{1}{\varepsilon_0}\sqrt{
    \frac{1}{N}\sum_{i=1}^N \left[\frac{1}{t}\int_0^t \varepsilon_i(t')dt'
    -\varepsilon_0\right]^2
    } \  \ . 
\end{equation}
$\Delta(t)$ represents deviation from equipartition of energy among particles.
At each time step $\varepsilon_i(t)$ varies from particle to particle.
A particle which initially has high energy
may give the energy  to particles which initially have low energy, 
and vice versa. Hence, in the course of time evolution,  
if the value %of time average
$\displaystyle \left(\frac{1}{t}\int_0^t \varepsilon_i(t')dt' - \varepsilon_0\right)$ 
decays  to  zero for most of the particles, 
we can say that  energy is well exchanged among particles.
On the other hand, $\Delta(t)$ remains constant 
if energy is not exchanged between particles
and every  $\varepsilon_i(t)$ is kept constant.
Thus decrease of $\Delta(t)$ indicates energy exchange among particles.

Fig.~\ref{fig:delta} shows temporal evolution of $\Delta(t)$ for the 
parameters~(\ref{eq:parameters}). 
Initial increase of $\Delta(t)$ represents that, instead of 
becoming equipartitioned,
differences of one-particle energy among particles are temporarily
enhanced. 
Let us turn our attention to  behavior of $\Delta(t)$ in later time.
We see that $\Delta(t)$ begins to decrease
from $t\sim 1500$, which is of the same order of the time
when the power-law fades away as seen in Fig.~\ref{fig:exponent-t}.
Hence we can say that the power-law structure relaxes as
energy is exchanged among particles.

\subsection{$\nu(\varepsilon)$  : cumulative energy distribution}
Now we investigate further the process of relaxation of power-law structure.
We have seen that the power-law structure begins to 
disappear as energy exchange
among particle begins, as indicated by the decrease of $\Delta(t)$.
If, through this relaxation,  energy distribution is kept constant,
this relaxation phase corresponds to ``{\it microscopic relaxation}''
discussed by Tsuchiya et al. ~\cite{sheet-tkg-1,sheet-tgk-2}.
If, on the other hand, energy distribution is also relaxed to that of
thermal equilibrium, then the phase corresponds to  ``{\it macroscopic 
relaxation}''. 

We measure  energy distribution by
$\displaystyle \nu(\varepsilon)$ defined as follows:
\begin{equation}
  \label{eq:nu-e}
  \nu(\varepsilon) \equiv
\frac{1}{N}\left(\mbox{number of particles whose energy $\varepsilon_i$ is 
    less than $\varepsilon$}\right)  \ ,
\end{equation}
where $N$ is the total number of particles.
Since $\displaystyle \nu(\varepsilon)$  is a cumulative distribution,
$\displaystyle \nu(\varepsilon)$ increases monotonically with respect to 
$\varepsilon$. If the graph of $\displaystyle \nu(\varepsilon)$ 
does not change with respect to time through relaxation, then the relaxation
is microscopic relaxation.

\begin{figure}[hbtp]
    \epsfxsize=6cm
    \epsfbox{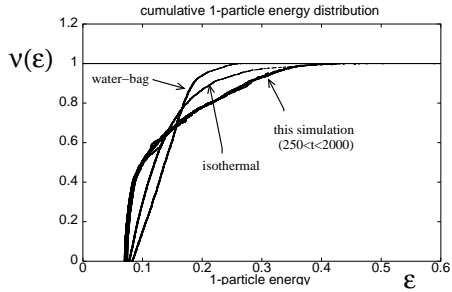}
  \caption{$\nu(\varepsilon)$ for $t=250,300,\cdots,2000$ }
  \label{fig:ne-relax}
\end{figure}

\begin{figure}[hbtp]
    \epsfxsize=6cm
    \epsfbox{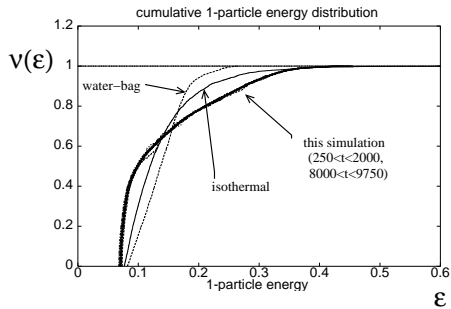}
  \caption{$\nu(\varepsilon)$ for 
    $t=250,300,\cdots,2000$, $8000,8050,\cdots,9750$ .}
  \label{fig:ne-relax2}
\end{figure}

Figs.~\ref{fig:ne-relax} and \ref{fig:ne-relax2} show
 $\nu(\varepsilon)$ for two temporal
intervals. Two well-known quasi-stationary
state, water-bag distribution and 
isothermal distribution~\cite{sheet-tkg-1,sheet-tgk-2}
are also shown as references.

Fig.~\ref{fig:ne-relax} represents $\nu(\varepsilon)$ for
$250<t<2000$, when the system has power-law correlation and 
the system is virialized (see Fig.~\ref{fig:vr}).
Please note that $\nu(\varepsilon)$ does not change for this time.

Fig.~\ref{fig:ne-relax2} represents $\nu(\varepsilon)$ for 
$8000 < t < 9750$ overlaid on the previous  figure. 
In this temporal interval
power-law correlation is lost (see Fig.~\ref{fig:exponent-t}).
It is clearly seen that the energy distribution remains the same
regardless of the existence of power-law correlation.
Thus we can say that the relaxation process of power-law structure
can be explained by the microscopic relaxation caused by energy exchange among particles
while keeping the global energy distribution fixed.

\section{Summary and discussions}
In this paper we numerically 
investigated the long-time behavior of power-law correlation
structure~\cite{hktk-1,hktk-2} 
in one-dimensional self-gravitating system. The power-law correlation 
exists even after the system is virialized. In this sense the structure
persists for long time. 
The lifetime of the structure  is, however, not infinite and 
it gradually fades away. By comparing the timescale of the lifetime 
with other time scales of relaxation we see that the decay of
the power-law structure is not caused by virialization process but
by energy exchange among particles.
The one-particle energy distribution $\nu(\varepsilon)$  is invariant through
the relaxation of the structure, hence it is the microscopic 
relaxation~\cite{sheet-tkg-1,sheet-tgk-2}.
The system is not stationary if we see it by the exponent $\alpha$ 
of correlation, but it is at the same time quasi-stationary if we see it
by the energy distribution $\nu(\varepsilon)$ after the system is
virialized.  

The fact that power-law correlation structure lasts long implies that 
the structure is important in the whole  dynamical evolution of the 
system and we may be able to observe such structure in other systems
even if the system is virialized.

Relation between the decay of power-law correlation and microscopic relaxation
has an important consequence. It is known that the microscopic relaxation 
time depends linearly on the system size
 $\displaystyle t_{micro}\propto N$~\cite{sheet-tgk-2}.
 Hence it is likely that
the average lifetime of the power-law correlation has similar dependence
on $N$, which is quite long for large $N$.

Let us now summarize schematically the entire life of the spatial 
power-law correlation structure in the model
 (\ref{eq:sheet-hamiltonian}) clarified through a series of our works.
\begin{enumerate}
\item 
From  initial condition with zero velocity dispersion~\cite{hktk-1}, \ 
spatial structure with power-law correlation emerges 
in small spatial scale~\cite{hktk-2}.
Typical example of such initial condition is $\forall i \  v_i = 0$ , 
$x_i = \mbox{random}$.

\item The power-law correlation 
structure grows up to large scale through hierarchical
clustering~\cite{hktk-2}.

\item The system is virialized. Power-law correlation structure persists.

\item Microscopic relaxation takes place and power-law 
correlation structure disappears. Energy distribution remains invariant.

\end{enumerate}

Related models with a uniform background or 
 friction~\cite{aurell-2001,miller-rouet-2001} are 
investigated from cosmological interest.
In those models similar fractal structure has been observed
and is more persistent.
Comparison with our results may be helpful %~\footnote{**** or ``give some hints''}
to understand mechanism by which the structure is retained in their models.

It should also be noted that the structure formation we have studied
have origin in dynamics.  It is interesting that,
instead of monotonously relaxing to thermal equilibrium 
the system, initially does not have spatial correlation, 
creates spatial correlation of power-law type through dynamics.
It seems difficult to explain it in terms of
thermodynamics or equilibrium statistical physics, since the structure appears
in states which are not thermally relaxed.  
Recently dynamical formation of spatial structure in 
long-range interacting systems is actively investigated.
Relation between our results and other examples of structure formation
with dynamical origin~\cite{tk-cluster,onion,dauxois-2000,barre-2001} will be interesting.

In this paper we have  focused on  typical behavior of the  system.
Several problems related to details of the behavior remain unsolved.
Among them there are, for example, 
initial condition dependence, system size dependence, and 
so on. These are beyond the scope of this short paper and  of 
important subjects of future study.

In this paper we have studied the behavior of the system until 
power-law structure disappears through energy exchange and energy
 distribution. It will be interesting to investigate whether there are
other relaxation processes effectively  working~\cite{lynden-bell-1967,rouet-feix-1999}, and the behavior of the system in the course of approach to 
equilibrium~\cite{sheet-tkg-1,sheet-tgk-2,yawn-miller-prl-1997,yawn-miller-pre-1997}.

The spontaneous emergence of power-law correlation 
we have studied so far is a novel phenomenon with rich implication. 
We hope our research serves as a foundation for future study,
both fundamental and applied.


\begin{thebibliography}{10}

\bibitem{hktk-1}
Hiroko Koyama and Tetsuro Konishi.
\newblock {\em Phys. Lett. A}, 279:226--230, 2001.

\bibitem{hktk-2}
Hiroko Koyama and Tetsuro Konishi.
\newblock astro-ph/0008507.





\bibitem{Miller-2}
B.N. Miller and C.J. Reidl Jr. 
\newblock {\em The Astrophysical Journal}, 348:203--211, 1990.


\bibitem{sheet-tkg-1}
T.~Tsuchiya, T.~Konishi, and N.~Gouda.
\newblock {\em Phys. Rev}, E 50:2607--2615, 1994.
\bibitem{sheet-tgk-2}
T.~Tsuchiya, N.~Gouda, and T.~Konishi.
\newblock {\em Phys. Rev}, E 53:2210--2216, 1996.

\bibitem{hohl-feix-1967}
Frank Hohl and Marc~R. Feix.
\newblock {\em The Astrophysical Journal}, 147:1164--1180, 1967.


\bibitem{rybivki-1971-apss}
George~B. Rybicki.
\newblock {\em Astrophysics and Space Science}, 14:56--72, 1971.

\bibitem{severne-1984-apss}
M.~Luwel, G.~Severne, and P.J. Rousseeuw.
\newblock {\em Astrophysics and Space Sciences}, 100:261--277, 1984.


\bibitem{yamashiro-gouda-sakagami}
T.~Yamashiro, N.~Gouda, and M.~Sakagami.
\newblock {\em Progress of Theoretical Physics}, 88:269--282, 1992.


\bibitem{tk-cluster}
T.~Konishi and K.~Kaneko.
\newblock {\em J. Phys.}, A 25:6283 -- 6296, 1992.

\bibitem{onion}
K.~Kaneko and T.~Konishi.
\newblock {\em Physica}, D 71:146--167, 1994.

\bibitem{aurell-2001}
E.~Aurell et al. .
\newblock {\em Physica}, D 148:272--288, 2001.

\bibitem{miller-rouet-2001}
B.~N.~Miller and J.~L.~Rouet.
\newblock astro-ph/0106117

\bibitem{dauxois-2000}
T.~Dauxois, P.~Holdsworth and S.~Ruffo.
\newblock {\em European Physical Journal} B 16: 659--667, 2000.

\bibitem{barre-2001}
J. Barr\'{e}, T.~Dauxois and S.~Ruffo.
\newblock {\em Physica} A 295:254--260, 2001

\bibitem{lynden-bell-1967}
D.~Lynden-Bell.
\newblock {\em Mon. Not. Roy. Astron. Soc. } 136:101--121 , 1967.

\bibitem{rouet-feix-1999}
J.~L.~Rouet and M.~R.~Feix.
\newblock {\em Phys. Rev.}E 59:73--83, 1999.  

\bibitem{yawn-miller-prl-1997}
K.~R.~Yawn and B.~N.~Miller.
\newblock {\em Phys. Rev. Lett.} 79:3561--3566 , 1997.

\bibitem{yawn-miller-pre-1997}
K.~R.~Yawn and B.~N.~Miller.
\newblock {\em Phys. Rev.} E 56:2429--2436 , 1997.


\end{thebibliography}
\end{document}